\input{aipcheck}

\documentclass[
    ,final            
  ]
  {aipproc}

\layoutstyle{6x9}

\usepackage{amsmath}
\newcommand{\nc}{\newcommand}
\nc{\lb}{\langle}
\nc{\rk}{\rangle}
\nc{\Blb}{\Big\langle}
\nc{\Brk}{\Big\rangle}

\nc{\mi}{\!\!\mid\!\!}
\nc{\ra}{\rightarrow}
\nc{\Ra}{\Rightarrow}
\nc {\cd}{\partial}
\nc {\sla}{\slashed}
\nc{\ro}{\mathrm}
\nc{\ca}{\mathcal}
\nc{\bo}{\mathbf}
\nc{\Tr}{\ro{Tr}\,}
\nc{\Str}{\ro{Str}}
\nc{\realtrace}{\ro{Re\; Tr}}
\nc{\maxrealtrace}{\ro{max\, Re\; Tr}}
\nc{\ud}{\ro{d}}
\nc{\nn}{\nonumber}

\nc{\pb}{\bar{\psi}}
\nc{\p}{\psi}
\nc{\Pb}{\bar{\Psi}}
\nc{\vp}{\vec{\pi}}
\nc{\vap}{\varphi}
\nc{\vt}{\vec{\tau}}
\nc{\si}{\sigma}
\nc{\Si}{\Sigma}
\nc{\g}{\gamma}
\nc{\G}{\Gamma}
\nc{\la}{\lambda}
\nc{\La}{\Lambda}
\nc{\ep}{\epsilon}
\nc{\de}{\delta}
\nc{\De}{\Delta}
\nc{\cL}{\ca{L}}
\nc{\cLe}{\ca{L}_{\ro{eff}}}
\nc {\ti}{\tilde}
\nc{\f}{\frac}
\nc{\da}{\dagger}
\nc{\SU}{\ro{SU}}
\nc{\om}{\omega}
\nc{\Om}{\Omega}

\nc{\darrow}{\stackrel{\leftrightarrow}{\cd}}
\nc{\darrows}{\stackrel{\leftrightarrow}{\sla{\cd}}}
\nc{\Darrows}{\stackrel{\leftrightarrow}{\sla{D}}}
\nc{\mr}{\stackrel{\circ}{m}_\rho}

\nc {\eqb}{\begin{equation}}
\nc {\eqe}{\end{equation}}
\nc {\eqab}{\begin{eqnarray}}
\nc {\eqae}{\end{eqnarray}}

\begin{document}

\title{Chiral effective field theory beyond the power-counting regime}


\author{Jonathan M. M. Hall}{
  address={Special Research Centre for the Subatomic Structure of
  Matter (CSSM), School of Chemistry \& Physics, University of
  Adelaide, SA 5005, Australia}
}

\author{Derek B. Leinweber}{
  address={Special Research Centre for the Subatomic Structure of
  Matter (CSSM), School of Chemistry \& Physics, University of
  Adelaide, SA 5005, Australia}
}

\author{Ross D. Young}{
  address={Special Research Centre for the Subatomic Structure of
  Matter (CSSM), School of Chemistry \& Physics, University of
  Adelaide, SA 5005, Australia}
}

\begin{abstract}
Novel techniques are presented, which identify the chiral 
power-counting regime (PCR), 
and realize the 
existence of an \emph{intrinsic energy scale} embedded 
in lattice QCD results that extend outside the PCR. 
The nucleon mass is considered as a 
 benchmark for illustrating this new 
approach. 
Using finite-range regularization, an optimal regularization scale  
can be extracted from lattice simulation 
results 
by analyzing the renormalization of the low energy coefficients. 
 The optimal scale allows a 
description of lattice simulation results that extend 
beyond the PCR  
by quantifying and thus handling any scheme-dependence.
 Preliminary results for the  
 nucleon magnetic moment are 
also examined, and a consistent 
optimal regularization scale is obtained. 
 This 
indicates the existence of an intrinsic scale corresponding to the finite 
size of the source of the pion cloud.
\end{abstract}

\maketitle


\section{Introduction}

 Chiral perturbation theory ($\chi$PT) provides a formal approach to
counting the powers of low energy momenta and quark masses, so that
an ordered expansion in 
the quark mass (or pion mass, through $m_q \propto m_\pi^2$)
 can be constructed.  $\chi$PT indicates that, in general, the most singular 
non-analytic contributions to hadron properties lie in the low order  
meson loop of the hadron.  For example, the leading non-analytic
behaviour of a baryon mass is proportional to $m_q^{3/2}$ or $m_\pi^3$. 
More generally, baryon masses can be written as an ordered expansion
in quark mass or $m_\pi^2$. 

 To establish a model-independent
framework in $\chi$PT, the expansion must display the properties of a
convergent series for the terms considered.  There is a power-counting
regime (PCR) where the quark mass is sufficiently small such that 
 higher order terms in
the expansion are negligible beyond the order calculated.  That is, 
within the
PCR, the truncation of the chiral expansion is reliable to a
prescribed precision \cite{Leinweber:2005xz,Young:2009ub}. 

The asymptotic nature of the chiral expansion places the focus on the
first few terms of the expansion.  
In chiral effective field theory ($\chi$EFT), 
calculations
 beyond one-loop are time-consuming 
 and there are no two-loop calculations which
incorporate the effects of placing a baryon in a finite volume. 
With
only a few terms of the expansion known for certain, knowledge of the PCR of
$\chi$EFT  is as important as knowledge of the expansion itself. 

Numerical simulations of QCD on a spacetime lattice are complemented
by $\chi$EFT through a model-independent formalism
for connecting lattice simulation results to the physical region. 
Simulations at finite volume and a variety of quark masses are related 
to the infinite-volume and physical quark masses through 
this formalism.  However, it is accurate only 
within the PCR of the truncated expansion.  
In particular, truncated expansions are 
typically applied to a wide range of quark (or pion) masses with 
little regard to a careful determination of the PCR. 

Through the consideration of a variety of renormalization schemes and
associated parameters, new techniques to identify the PCR 
have been established in Ref \cite{Hall:2010ai}. 
%
Consider the nucleon mass: a benchmark for illustrating this
 approach.

\section{Effective field theory for nucleons}
Using the standard Gell-Mann--Oakes--Renner relation connecting
quark and pion masses, $m_q \propto m_\pi^2$ \cite{GellMann:1968rz},
the formal chiral expansion of the nucleon mass can be written as the sum of 
a polynomial expansion in $m_\pi^2$ and the meson-loop integral
contributions: 
%
\eqb
M_N 
= \{a_0 + a_2 m_\pi^2 + a_4 m_\pi^4 + \ca{O}(m_\pi^6)\} 
+ \Si_N + \Si_\De + \Si_{tad}\,.
\label{eqn:mNresid}
\eqe

The pion cloud corrections are considered in the heavy-baryon limit,
with loop integrals, $\Si_N$, $\Si_\De$ and $\Si_{tad}$. 
The coefficients $a_i$ of the analytic polynomial 
in Eq.~(\ref{eqn:mNresid}) are related to the
low energy constants of $\chi$PT.  In this investigation, they will be
determined by fitting to lattice simulation results.  These
coefficients will be referred to 
as the \emph{residual series} coefficients.  These bare
coefficients undergo renormalization due to contributions from the
loop integrals $\Si_N$, $\Si_\De$ and $\Si_{tad}$. 

Under the most general considerations, each loop integral, when
evaluated, produces an analytic polynomial in $m_\pi^2$ and
non-analytic terms. Using finite-range 
regularization, expanding out each integral yields:
\begin{align}
\label{eqn:NN}
\Si_N &= \f{\chi_N}{2\pi^2}\int\!\!\ud^3\! k
\f{k^2 u^2(k^2\,;\La)}{\om^2(k)}
%
= b_0^N + b_2^N m_\pi^2 + \chi_N m_\pi^3 + b_4^N m_\pi^4 +
 \ca{O}(m_\pi^5)\,,\\
\label{eqn:ND}
\Si_\De &= \f{\chi_\De}{2\pi^2}\int\!\!\ud^3\! k
\f{k^2 u^2(k^2\,;\La)}{\om(k)\left(\De + \om(k)\right)}
= b_0^\De + b_2^\De m_\pi^2  + b_4^\De m_\pi^4 
+\f{3}{4\pi\De} 
\chi_\De m_\pi^4\,\log\f{m_\pi}{\mu} + \ca{O}(m_\pi^5)\,,\\
\label{eqn:tad}
\Si_{tad} &= c_2 m_\pi^2\Big(
\chi_t\f{1}{4\pi}\int\!\!\ud^3\! k
\f{2 u^2(k^2\,;\La)}{\om(k)} - b_0^{t}\!\Big)
= c_2\!\Big(\!b_2^{t} m_\pi^2  +  b_4^{t} m_\pi^4 + \chi_t
 m_\pi^4\,\log\f{m_\pi}{\mu} + \ca{O}(m_\pi^5)\!\!\Big),
\end{align}
where $u(k^2\,;\La)$ is a regulator function described below. 
These integrals are expressed in terms of the pion energy $\om(k) =
\sqrt{k^2 + m_\pi^2}$ and 
the delta-nucleon mass splitting $\De$.  $\chi_N$, $\chi_\Delta$ and $\chi_t$
denote the model-independent chiral coefficients of the terms that are
non-analytic in the quark mass.  
The $b_i$ coefficients are
renormalization-scheme dependent, as are the $a_i$ coefficients.

The process of renormalization in finite-range regularized
 $\chi$EFT proceeds by combining
the renormalization-scheme dependent coefficients to provide the
physical low energy coefficients, which are denoted as $c_i$.  Thus, the
nucleon mass expansion takes on the standard form:
%
\eqb
M_N 
= c_0 + c_2 m_\pi^2 + \chi_N m_\pi^3 +
c_4 m_\pi^4 
+ \!\left(\!\!-\f{3}{4\pi\De}\chi_\De + c_2\chi_t
 \!\right)\!m_\pi^4\log\f{m_\pi}{\mu} + \ca{O}(m_\pi^5)\,.
\label{eqn:mNexpansion}
\eqe
%
By comparing Eqs.~(\ref{eqn:mNresid}) and (\ref{eqn:mNexpansion}),
 the following renormalization procedure is obtained:
\eqab
\label{eqn:c0norm}
c_0 &=& a_0 + b_0^N + b_0^\De\,,\\
\label{eqn:c2norm}
c_2 &=& a_2 + b_2^N + b_2^\De + c_2 b_2^{t}\,,\\
\label{eqn:c4norm}
c_4 &=& a_4 + b_4^N + b_4^\De + c_2 b_4^{t}\,, \mbox{\,\,etc.}
\eqae
The coefficients $c_i$ are scheme-independent quantities, and
 this property will be demonstrated when determined within the PCR.

 For the purposes of this investigation, 
 the regulator function $u(k\,;\La)$ is chosen to be a dipole form:
\eqb
u(k\,;\La) = {\left(1 + \f{k^2}{\La^2}\right)}^{-2}.
\eqe
More 
 extensive discussions of different finite-range regulator 
 functional forms can be found in 
Refs \cite{Leinweber:2005xz,Leinweber:2003dg,Bernard:2003rp,Leinweber:2005cm}. 
An analysis of results for the nucleon mass 
for a family of dipole-like forms 
is reported in Ref \cite{Hall:2010ai}.

 In order to accommodate the effect of the finite 
volume, the continuous loop integrals 
are transformed into a sum over
discrete momentum values $k_i=n_i\,2\pi/L$, for integers $n_i$.  
The difference between a loop sum and its
corresponding loop integral is the finite-volume correction, which
should vanish for all integrals as $m_\pi L$ becomes large
\cite{Beane:2004tw}. 
%

\section{The intrinsic scale}

\begin{figure}
\begin{minipage}[b]{0.5\linewidth} 
\centering
\includegraphics[height=1.0\hsize,angle=90]{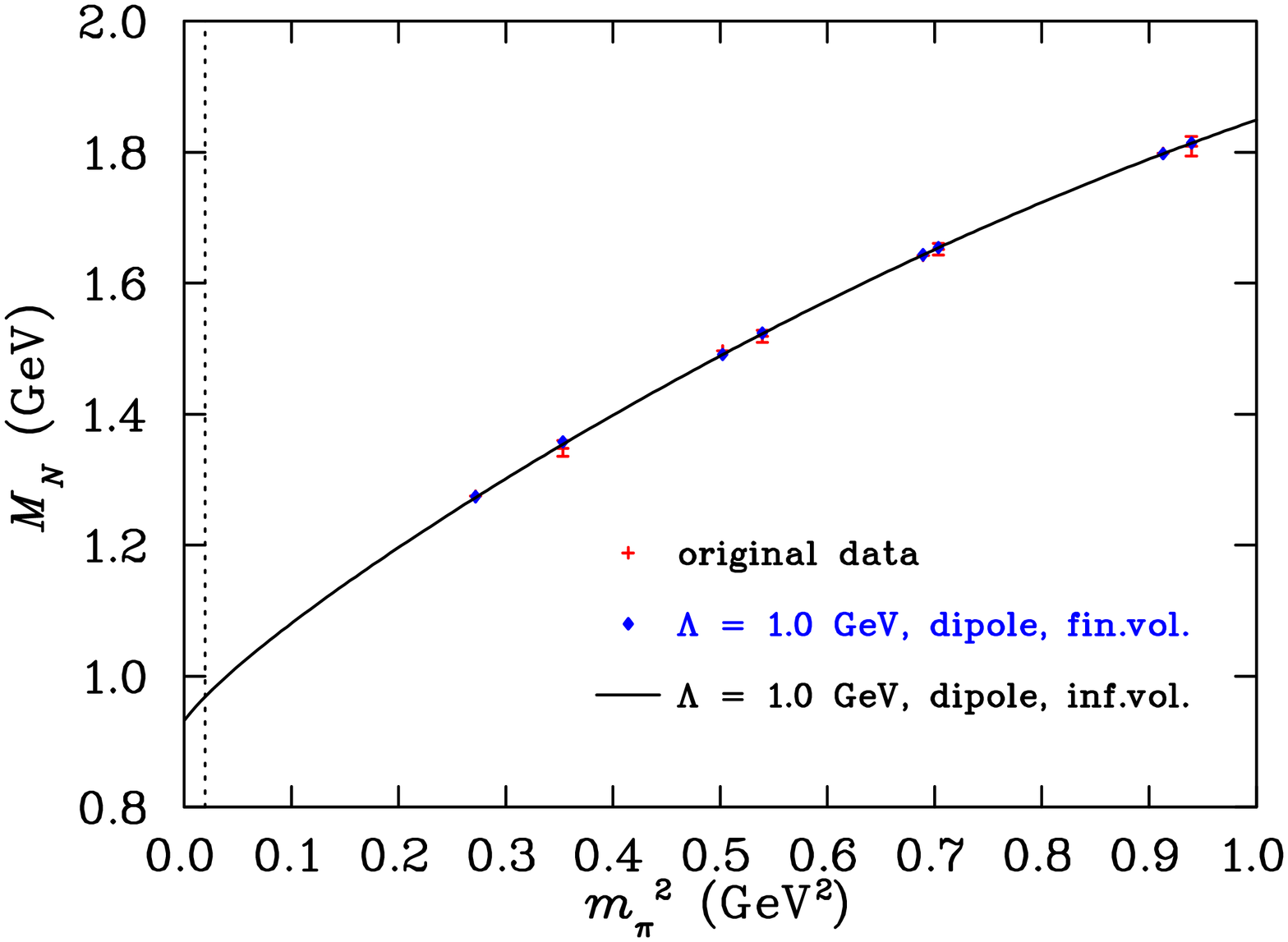}
\vspace{-6mm}
\label{fig:YoungExt}
\end{minipage}
\hspace{2mm}
\begin{minipage}[b]{0.5\linewidth} 
\centering
\includegraphics[height=1.03\hsize,angle=90]{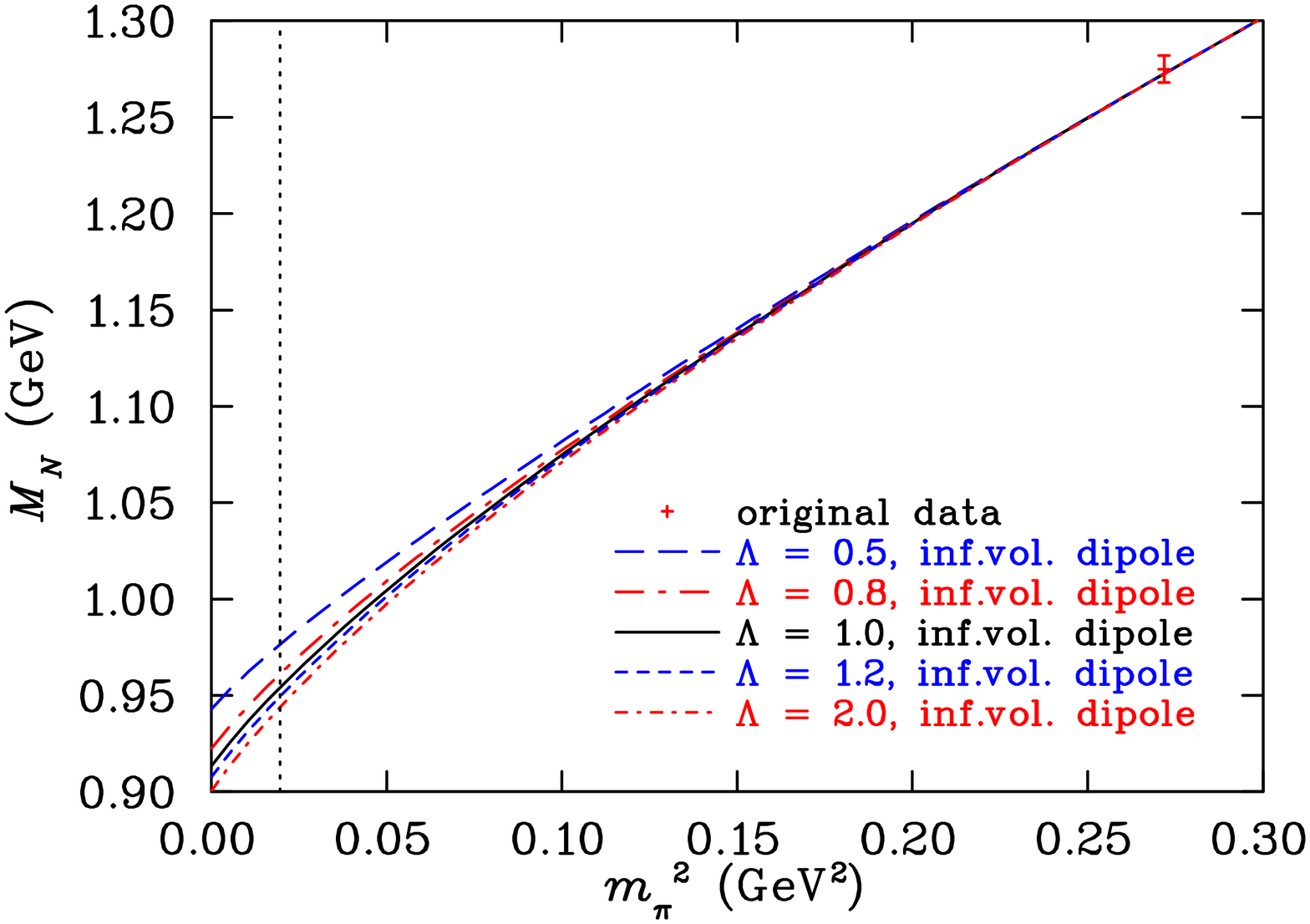}
\vspace{-6mm}
\caption{\footnotesize{\textit{Left}: example dipole extrapolation based on CP-PACS data \cite{AliKhan:2001tx}, box size: $2.3-2.8$ fm. \textit{Right}: close zoom of the regulator-dependence of the extrapolation.}}
\label{fig:YoungExtmulti}
\end{minipage}
\end{figure}

The $\chi$EFT extrapolation scheme 
 will be used in conjunction with 
lattice QCD data from CP-PACS \cite{AliKhan:2001tx} and  
PACS-CS \cite{Aoki:2008sm} to predict the nucleon mass $M_N$
for any value of $m_\pi^2$. 
When using lattice QCD results 
beyond the PCR, there is a scheme-dependence in the extrapolation. That is, 
 the result of an extrapolation will be dependent on the choice of 
regulator.
To demonstrate this, consider the infinite-volume extrapolation of the
CP-PACS data as the regularization scale is changed.
Figure \ref{fig:YoungExtmulti} shows that the extrapolations 
 diverge in the chiral regime, 
where the non-analytic behaviour of the loop integrals dominate. 
%
\subsection{Identification of the intrinsic scale: a toy model}

Is there a `best' regularization scale indicated by lattice QCD data? 
By constructing a toy model, 
a method is developed to identify an in-built scale by analyzing the 
low energy coefficients.  Ideal pseudodata is created with 
  a known scale $\Lambda_c$, and an 
  examination of the renormalization flow of the low energy 
 coefficients $c_i$ allows one to recover the value of the scale $\Lambda_c$. 
%

\begin{figure}[tp]
\begin{minipage}[b]{0.5\linewidth} 
\centering
\includegraphics[height=1.03\hsize,angle=90]{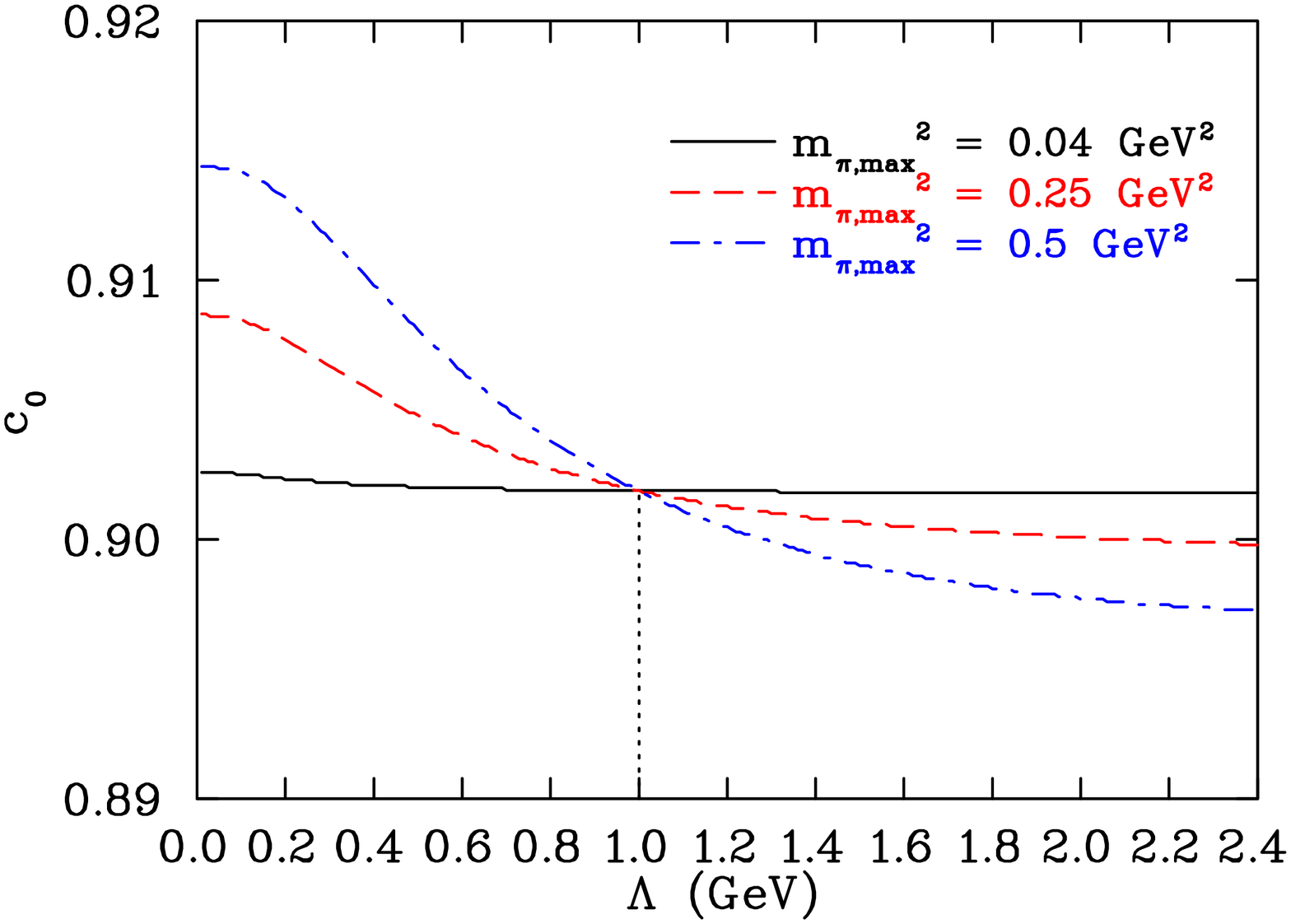}
\end{minipage}
\hspace{2mm}
\begin{minipage}[b]{0.5\linewidth} 
\centering
\vspace{4mm}
\includegraphics[height=1.0\hsize,angle=90]{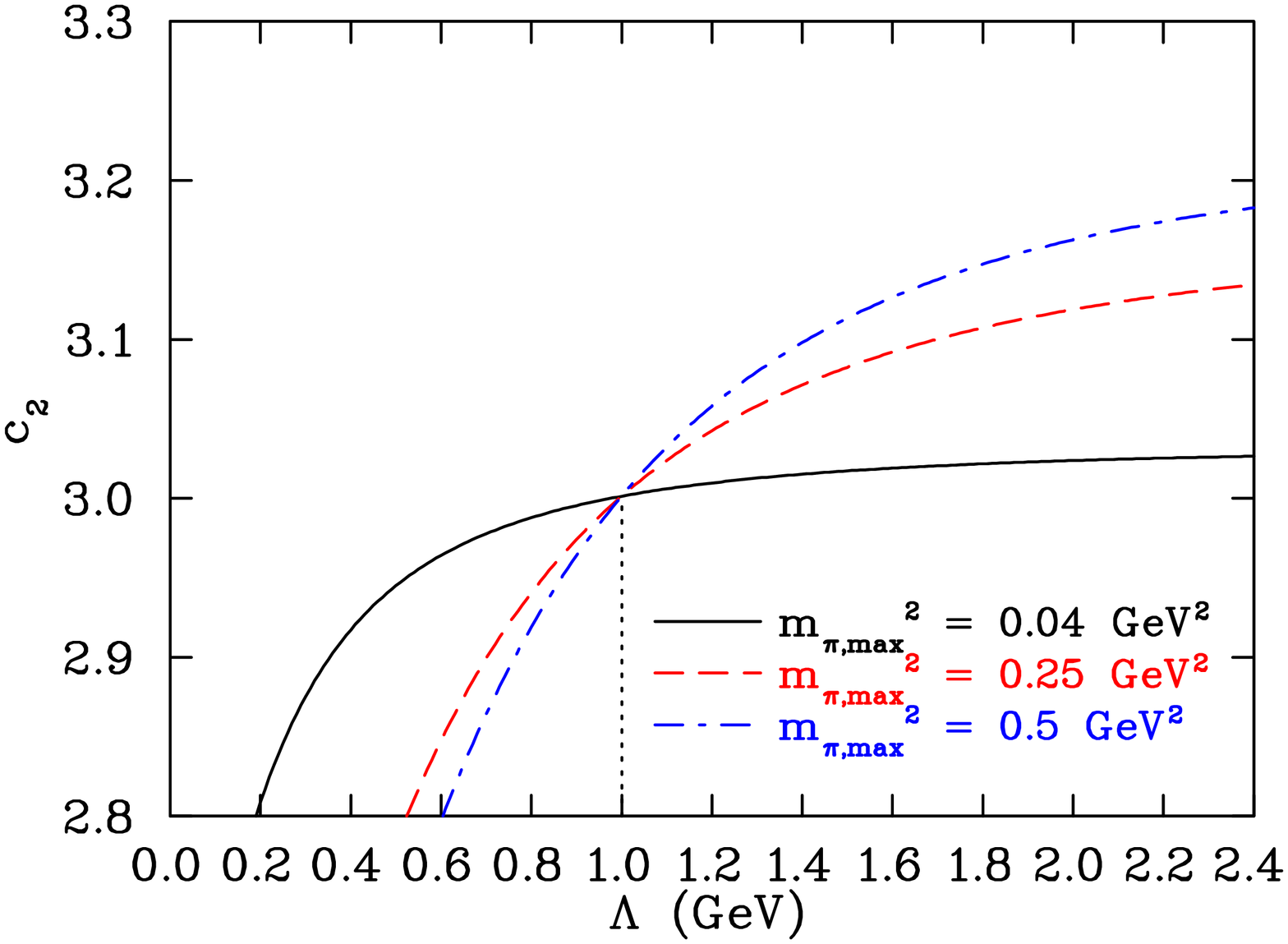}
\vspace{-6mm}
\caption{\footnotesize{ \textit{Left}: Behaviour of $c_0$ vs.\ $\La$.
\textit{Right}: Behaviour of $c_2$ vs.\ $\La$. 
Both plots are based on infinite-volume pseudodata created with a dipole regulator at $\La_\ro{c} = 1.0$ GeV (based on lightest four data points from PACS-CS). Each curve follows from fits of Eq.~(\ref{eqn:mNresid}) to pseudodata, with a different upper value of pion mass $m_{\pi,\ro{max}}^2$ considered. }}
\label{fig:pdata}
\end{minipage}
\end{figure}

%
Based on PACS-CS data, 
a dense and
precise set of pseudodata 
is generated at infinite volume and $\La_c = 1.0$ GeV, entirely within the PCR.
 As the pseudodata are extended beyond the PCR 
(by increasing $m_{\pi,\ro{max}}^2$),
 a scale-dependence in the renormalization 
is realized, as shown for $c_0$ and $c_2$ in Figure \ref{fig:pdata}.

However, the correct values of $c_0$ and $c_2$ (known by construction) 
are recovered at exactly $\La = \La_\ro{c}$, where the renormalization 
is independent of the range of $m_\pi^2$ used. 
It is \emph{not} correct in general 
to assume that the 
 asymptotic value of $c_i$ at large $\La$ is the best estimate of the true 
$c_i$. Note that the intersection point is at the same value of $\La$ for both 
 $c_0$ and $c_2$. This indicates that the optimal regulator for the pseudodata 
is the value $\La = \La_\ro{c}$.

\subsection{Lattice simulation results}

In the toy model, a method was developed for extracting the intrinsic 
scale from some ideal pseudodata. 
 This leads one to ponder if actual lattice simulation results contain 
an intrinsic scale. Since lattice data usable for an extrapolation 
 generally extend outside the PCR, 
  an optimal regularization scale may be recovered.   
 Renormalization flow curves can be constructed for lattice simulation results 
 by constraining the 
 fit window with varying values of 
 $m_{\pi,\ro{max}}^2$, sequentially including data 
points extending further outside the PCR.

\begin{figure}
\begin{minipage}[b]{0.5\linewidth}
\centering
\includegraphics[height=1.0\hsize,angle=90]{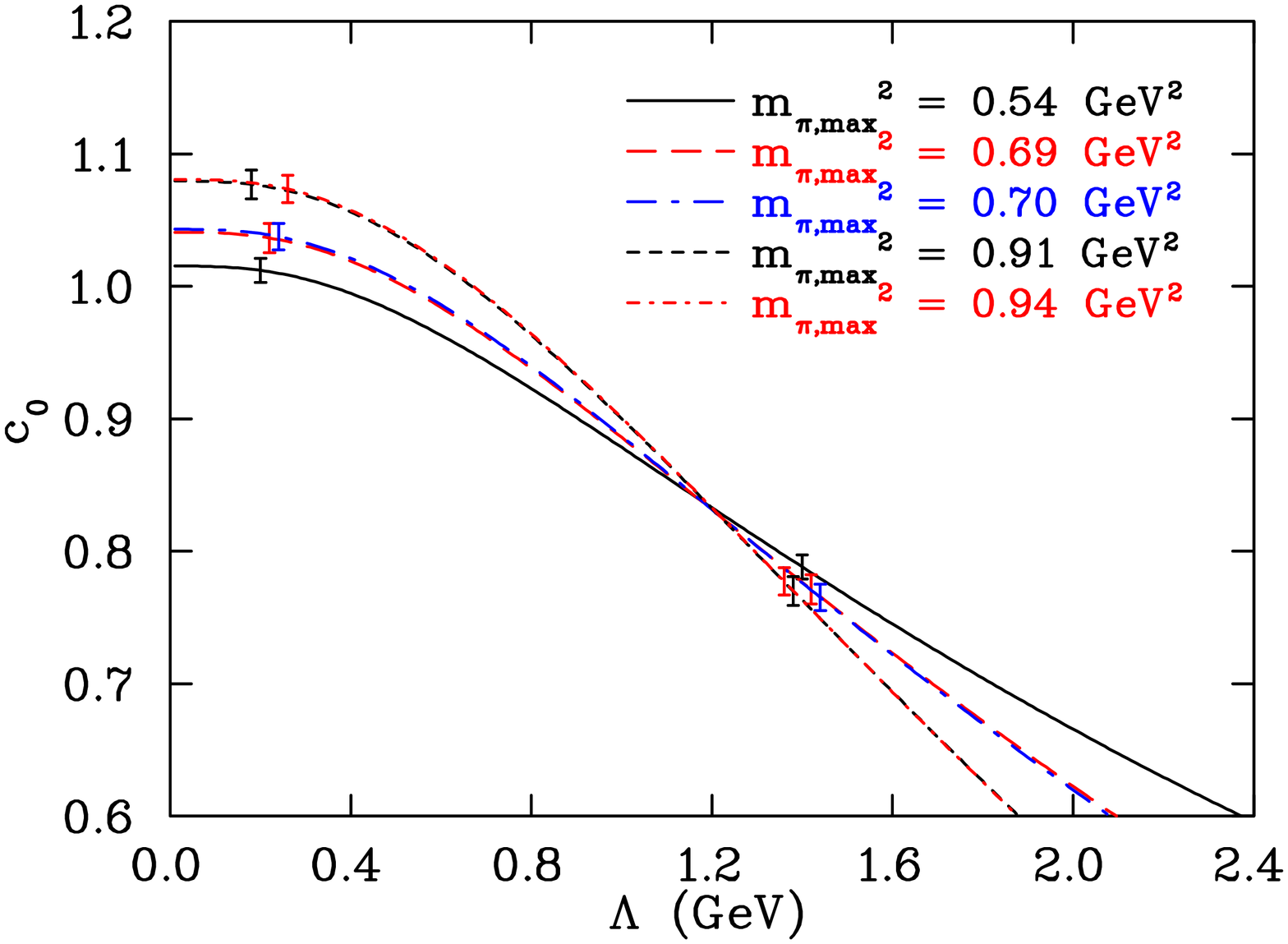}
\vspace{-6mm}
\end{minipage}
\begin{minipage}[b]{0.5\linewidth}
\centering
\includegraphics[height=1.0\hsize,angle=90]{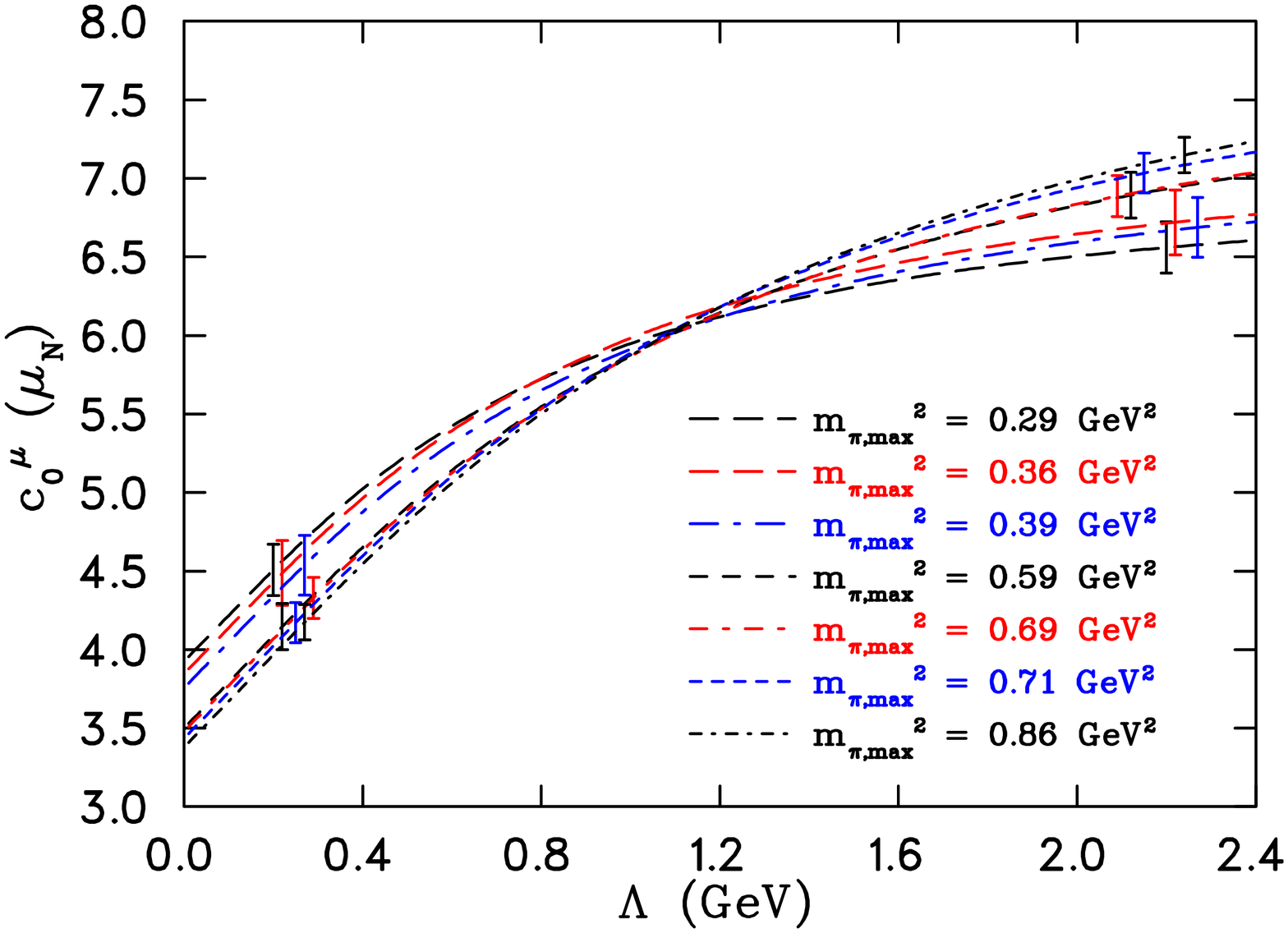}
\vspace{-6mm}
\caption{\footnotesize{\textit{Left}: Behaviour of $c_0$ vs.\ $\La$, based on CP-PACS data, to order $\ca{O}(m_\pi^3)$. \textit{Right}: Behaviour of $c_0^\mu$ vs.\ $\La$, based on QCDSF data, to order $\ca{O}(m_\pi)$. In both plots a dipole regulator is used. A few points are selected to indicate the general size of the statistical error bars.}}
\end{minipage}
\label{fig:c0actual}
\end{figure}

The results for the renormalization 
of the low energy coefficients $c_0$, the nucleon mass in the chiral limit, and 
   $c_0^\mu$, the magnetic moment of the nucleon in the chiral limit, 
are shown in Figure \ref{fig:c0actual}. Lattice simulation results 
from CP-PACS are used to analyze the nucleon mass, and preliminary 
results from QCDSF are used to 
analyze the magnetic moment. 
 Similar to the pseudodata case, a greater degree of regulator-dependence 
is observed as more data are included in the fit. 
There is also a reasonably well defined regularization scale  
at which the renormalization of $c_0$ is insensitive to the truncation 
of the data. This indicates that there exists an optimal regularization scale 
embedded 
in the lattice simulation results themselves. 

It is also intriguing that the 
 optimal regularization scale 
obtained from independent lattice results for different 
observables is approximately the same value 
for the same regulator functional form. This suggests that the 
 optimal regularization scale 
indicates the existence of an intrinsic scale in the nucleon-pion interaction.

\begin{figure}
\begin{minipage}[b]{0.5\linewidth}
\centering
\includegraphics[height=1.0\hsize,angle=90]{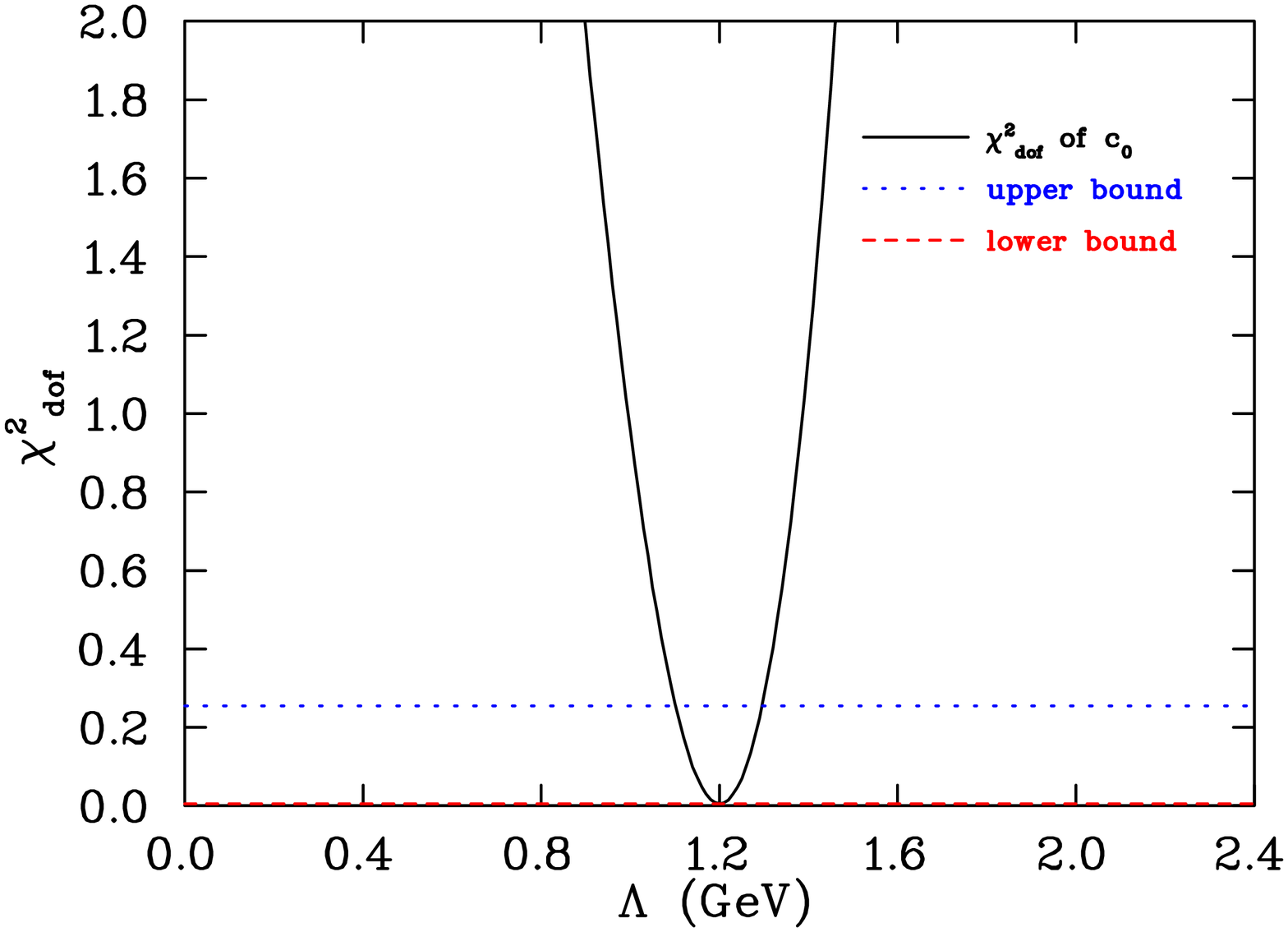}
\vspace{-6mm}
\end{minipage}
\begin{minipage}[b]{0.5\linewidth}
\centering
\includegraphics[height=1.0\hsize,angle=90]{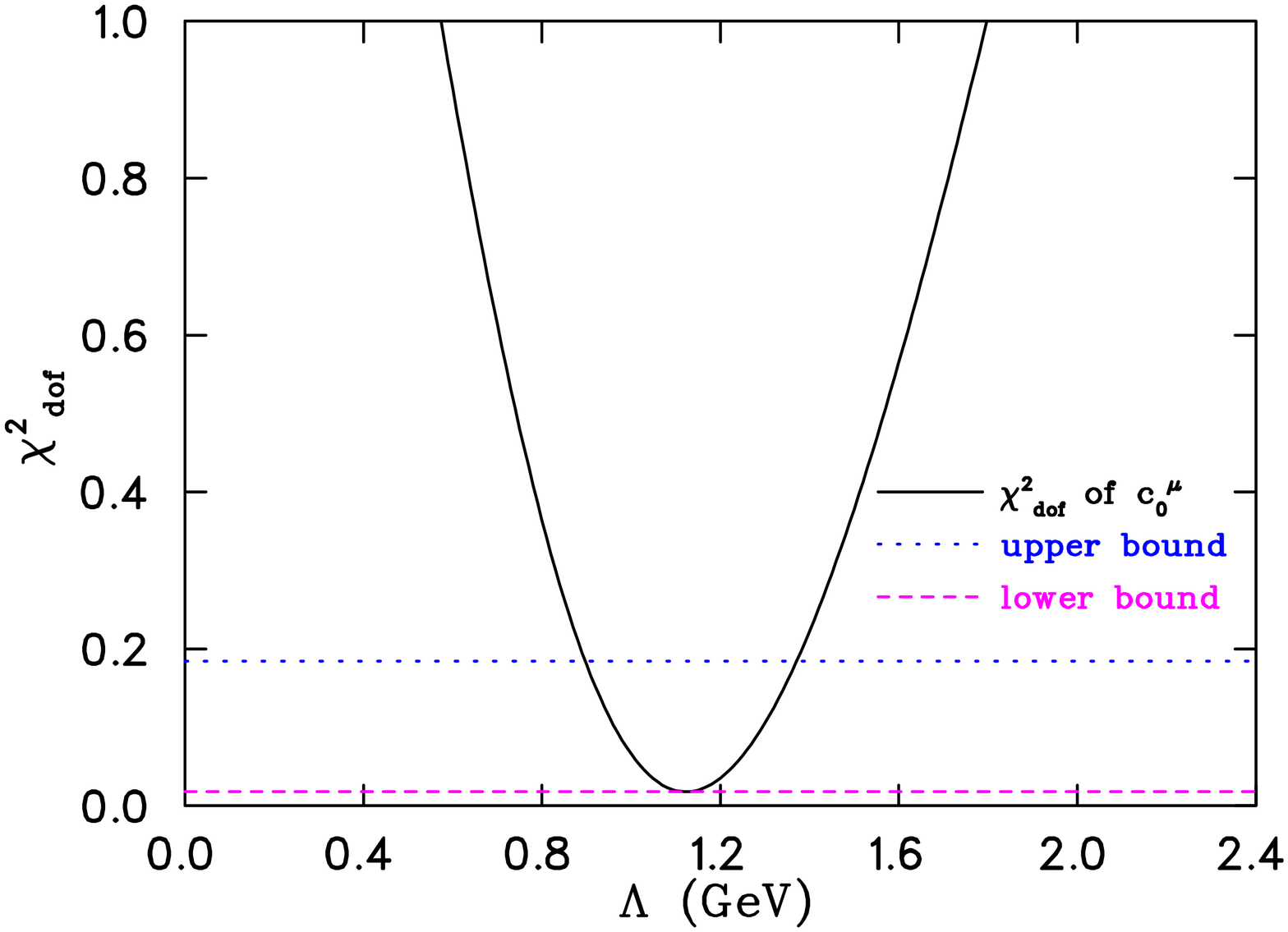}
\vspace{-7mm}
\caption{\footnotesize{\textit{Left}: Behaviour of $\chi^2_{dof}$ vs.\ $\La$, for the renormalization flow of $c_0$ to order $\ca{O}(m_\pi^3)$, based on CP-PACS data. \textit{Right}: Behaviour of $\chi^2_{dof}$ vs.\ $\La$, for the renormalization flow of $c_0^\mu$ to order $\ca{O}(m_\pi^3)$, based on QCDSF data. In both plots a dipole regulator is used.}}
\end{minipage}
\label{fig:c0chisqdofactual}
\end{figure}

A measure of the systematic uncertainty in the optimal regularization scale, 
evident in Figure \ref{fig:c0actual}, can be obtained  
 by constructing a $\chi^2$-style analysis.
By plotting $\chi^2$ against the regularization scale $\La$, 
 a measure of the spread of the 
renormalization 
flow curves can be calculated, and the intersection point obtained. 
The corresponding $\chi^2$ (per degree of freedom) plots are shown in 
Figure \ref{fig:c0chisqdofactual}.

\section{Summary}
\vspace{-2.5mm}
Using finite-range regularized chiral effective field theory, 
a technique for obtaining an optimal regularization scale 
from lattice results   
was demonstrated 
for pseudodata, in addition to actual 
simulation results for the nucleon mass and  
 simulation results for the nucleon magnetic moment. 
By analyzing the 
renormalization flow of the low energy coefficients with respect 
to $\La$, a step-wise extension of
 the data beyond the PCR 
 allowed the identification of an optimal regularization scale.
 This scale of approximately $1.2$ GeV 
is the value at which the renormalization is
  least sensitive to truncation of the lattice 
data.
 Lattice simulation results for the nucleon mass and nucleon magnetic moment 
yielded consistent values for the optimal scale. 
Thus an intrinsic scale was uncovered, which characterizes 
the energy scale of the pion dressing of the nucleon.


\vspace{-3.5mm}
\begin{theacknowledgments}
This research 
is supported by the Australian Research Council.
\end{theacknowledgments}

\vspace{-3.5mm}
\bibliographystyle{aipproc}   

\bibliography{cairnsref}

\begin{thebibliography}{10}
\expandafter\ifx\csname natexlab\endcsname\relax\def\natexlab#1{#1}\fi
\providecommand{\enquote}[1]{``#1''}
\expandafter\ifx\csname url\endcsname\relax
  \def\url#1{\texttt{#1}}\fi
\expandafter\ifx\csname urlprefix\endcsname\relax\def\urlprefix{URL }\fi
\providecommand{\eprint}[2][]{\url{#2}}

\bibitem[Leinweber et~al.(2005)]{Leinweber:2005xz}
D.~B. Leinweber, A.~W. Thomas, and R.~D. Young, \emph{Nucl. Phys.}
  \textbf{A755}, 59--70 (2005), \eprint{hep-lat/0501028}.

\bibitem[Young et~al.(2009)]{Young:2009ub}
R.~D. Young, J.~M.~M. Hall, and D.~B. Leinweber  (2009), \eprint{0907.0408}.

\bibitem[Hall et~al.(2010)]{Hall:2010ai}
J.~M.~M. Hall, D.~B. Leinweber, and R.~D. Young, \emph{Phys. Rev.}
  \textbf{D82}, 034010 (2010), \eprint{1002.4924}.

\bibitem[Gell-Mann et~al.(1968)]{GellMann:1968rz}
M.~Gell-Mann, R.~J. Oakes, and B.~Renner, \emph{Phys. Rev.} \textbf{175},
  2195--2199 (1968).

\bibitem[Leinweber et~al.(2004)]{Leinweber:2003dg}
D.~B. Leinweber, A.~W. Thomas, and R.~D. Young, \emph{Phys. Rev. Lett.}
  \textbf{92}, 242002 (2004), \eprint{hep-lat/0302020}.

\bibitem[Bernard et~al.(2004)]{Bernard:2003rp}
V.~Bernard, T.~R. Hemmert, and U.-G. Meissner, \emph{Nucl. Phys.}
  \textbf{A732}, 149--170 (2004), \eprint{hep-ph/0307115}.

\bibitem[Leinweber et~al.(2006)]{Leinweber:2005cm}
D.~B. Leinweber, A.~W. Thomas, and R.~D. Young, \emph{PoS} \textbf{LAT2005},
  048 (2006), \eprint{hep-lat/0510070}.

\bibitem[Beane(2004)]{Beane:2004tw}
S.~R. Beane, \emph{Phys. Rev.} \textbf{D70}, 034507 (2004),
  \eprint{hep-lat/0403015}.

\bibitem[Ali~Khan et~al.(2002)]{AliKhan:2001tx}
A.~Ali~Khan, et~al., \emph{Phys. Rev.} \textbf{D65}, 054505 (2002),
  \eprint{hep-lat/0105015}.

\bibitem[Aoki et~al.(2009)]{Aoki:2008sm}
S.~Aoki, et~al., \emph{Phys.Rev.} \textbf{D79}, 034503 (2009),
  \eprint{0807.1661}.

\end{thebibliography}

\end{document}